\documentclass[graybox]{svmult}

\usepackage{type1cm}        
\usepackage{makeidx}         
\usepackage{graphicx}        
\usepackage{multicol}        
\usepackage[bottom]{footmisc}
\usepackage{tabularx}

\usepackage{newtxtext}       %
\usepackage[varvw]{newtxmath}       

\makeindex             

\usepackage{cases}

\def\mearth{\text{M}_\text{E}}
\def\mcore{M_\text{core}}

\def\menv{M_\text{env}}

\def\sigmag{\Sigma_g}

\def\omegak{\Omega_{\text{K}}}

\def\st{\text{St}}
\def\rhill{R_\text{H}}

\def\mdotplan{\dot{M}_\text{plan}}

\begin{document}

\title*{The formation of planetary systems: physics, populations, and architectures}
\author{Andrin Kessler \orcidID{0000-0002-4475-0685}, Jesse Weder \orcidID{0000-0001-8270-1756}, Jesse Polman \orcidID{0009-0009-8749-9513}, Nicolas Kaufmann \orcidID{0009-0005-4389-2947}, Jeanne Davoult \orcidID{0000-0002-6177-2085}, Alexandre Emsenhuber \orcidID{0000-0002-8811-1914}, Yann Alibert \orcidID{0000-0002-4644-8818}, and Christoph Mordasini\orcidID{0000-0002-1013-2811}}

\authorrunning{Kessler et al.}

\institute{Andrin Kessler \at Space Research \& Planetary Sciences,  University of Bern, Gesellschaftsstrasse 6, 3012 Bern, Switzerland, \email{andrin.kessler@unibe.ch} 
\and Jesse Weder \at Space Research \& Planetary Sciences,  University of Bern, Gesellschaftsstrasse 6, 3012 Bern, Switzerland, \email{jesse.weder@unibe.ch}
\and Jesse Polman \at Space Research \& Planetary Sciences,  University of Bern, Gesellschaftsstrasse 6, 3012 Bern, Switzerland, \email{jesse.polman@unibe.ch}
\and Nicolas Kaufmann \at University Observatory Munich, Scheinerstr. 1, 81679 München, Germany, \email{N.Kaufmann@lmu.de}
\and Jeanne Davoult \at Institut für Planetenforschung, German Aerospace Center (DLR), Rutherfordstraße 2, 12489 Berlin, Germany, \email{jeanne.davoult@dlr.de}
\and Alexandre Emsenhuber \at Space Research \& Planetary Sciences,  University of Bern, Gesellschaftsstrasse 6, 3012 Bern, Switzerland, \email{alexandre.emsenhuber@unibe.ch}
\and Yann Alibert \at Space Research \& Planetary Sciences,  University of Bern, Gesellschaftsstrasse 6, 3012 Bern, Switzerland, \email{yann.alibert@unibe.ch}
\and Christoph Mordasini, \at Space Research \& Planetary Sciences,  University of Bern, Gesellschaftsstrasse 6, 3012 Bern, Switzerland, \email{christoph.mordasini@unibe.ch}}

\maketitle

\abstract{In this chapter we review the progresses made in global theoretical models of planetary system formation in the last decade using the example of the state-of-the-art planetary system formation framework known as the Bern Model that has been continuously developed since before the beginning of the NCCR PlanetS. We highlight several major developments and applications that have since been implemented, reflecting important recent advancements of planet formation theory overall, such as  MHD wind-driven disk evolution,  planetesimal evolution including fragmentation, dust evolution and pebble accretion, formation of planets in structured disks,  interior structure models allowing for compositional gradients, as well as the analysis of the emerging planetary system architectures and the identification of different classes of architectures. We discuss how these new models impact the formation and evolution process and translate into different populations of planets and planetary systems.
{We also discuss the major strengths of the Bern Model, including successful predictions of the break in the planetary mass function at 30 $M_\oplus$, the prevalence of low-mass planets, the radius pile-up around 1 $R_\textrm{Jupiter}$, and the evaporation valley, with the recent New Generation Planetary Population Synthesis models with 100 seeds per disk providing quantitive matches to many RV-survey and Kepler diagnostics. This includes key characteristics of planetary system architectures. We also highlight the limitations of the Generation 3 Bern model, some of them were addressed during the course of the NCCR PlanetS: the inclusion of the early phases of planet formation from dust to planetesimals, the hybrid pebble-planetesimals accretion of solids, the simplified interior structure models, a general reliance on simplified parametrizations that may not encapsulate the full complexity of physical processes, and computational constraints including short (hundred Myr) N-body integration times.}}

\section{Introduction}
\label{sec: introduction}

In this chapter we review the progress made in global theoretical models of planet formation in the core accretion paradigm in the last decade. In contrast to the direct formation via gravitational instability in massive protoplanetary disks discussed in the previous chapter, here, the dust component of the protoplanetary disk first aggregates into the planetary cores, before the most massive cores are able to bind a gaseous envelope.\\

Planet formation is inherently stochastic, involving a large number of interconnected processes. The aim of global models of planet formation \cite{mordasiniGlobalModelsPlanet2015,RaymondMorbidelli2022} is to combine all relevant physical process into a self-consistent model -- from protoplanetary disk to planetary system. This allows to study the importance and interplay of all included processes which are often studied in isolation \cite{drazkowskaPlanetFormationTheory2022a}. However, it is not straightforward to link initial conditions to the characteristics of the emerging planetary systems. Therefore, it is unclear whether a well-fitting model describing a single planetary system, e.g. the Solar System, necessarily contains the relevant physical processes or whether it is just a randomly fitting outcome of incomplete physics. A way to overcome this problem is to test planet formation theory on a statistical level by comparing synthetic populations of planetary systems, emerging from global models, with the exoplanet population. In order to obtain a meaningful synthetic population, initial condition parameters of the protoplanetary disk must be sampled from distributions that reflect observations. The comparison and detailed study of formation pathways can hint at incomplete or inaccurate physics, which helps with identifying the specific physical processes that are most important to include or understand better in order to progress planet formation theory efficiently \cite{MordasiniBurn2024RvMG}.

Planetary population synthesis \cite{benzPlanetPopulationSynthesis2014,Mordasini2018haex,burnPlanetaryPopulationSynthesis2024} requires a global model in the sense that it contains the structure and evolution of the protoplanetary disk, accretion processes, planetary structure, planet-planet, as well as planet-disk interactions. Furthermore, the model must be simple enough to remain computationally efficient, allowing it to produce large planetary populations, given that a single population can contain up to a hundred thousand (proto)planets \cite{emsenhuberNewGenerationPlanetary2021}. Although there are significant observational uncertainties associated with the initial conditions of protoplanetary disks, population synthesis has revealed and contributed to the understanding of several key planet formation processes. For instance, early syntheses have identified that planets most likely do not form at the same location (in-situ) but are subject to orbital migration \cite{idaDeterministicModelPlanetary2008, mordasiniExtrasolarPlanetPopulation2009a}. This lead to the increasingly in-depth modelling of planetary migration in dedicated studies \cite{massetSaturatedTorqueFormula2010,kleyPlanetMigrationThreedimensional2009,paardekooperTorqueFormulaNonisothermal2010} which were in turn again incorporated in improved global models \cite{dittkristImpactsPlanetMigration2014}. Similarly, grain dynamics and opacities of planetary atmospheres have been shown to play a critical role in planet formation \cite{mordasiniGrainOpacityBulk2014} which has subsequently spawned more detailed models of dust dynamics \cite{ormelAtmosphericStructureEquation2014,mordasiniGrainOpacityBulk2014a,brouwersHowPlanetsGrow2020}. Examples such as these highlight the productive synergy between global and specialised models.\\

In Sect.~\ref{sec: planetary formation and evolution framework}, we introduce the global planet formation and evolution model that has been developed significantly and applied extensively, mainly at the University of Bern but also several partner institutes, during the period of the NCCR PlanetS. We give an overview of the physical processes that are considered as described in detail in Emsenhuber et al. 2021 \cite{emsenhuberNewGenerationPlanetary2021} in Sect.~\ref{sec: model description} and showcase the effect of orbital migration as one of the key processes studied in the last decade in Sect.~\ref{sec: Planetary migration on a population level}. In Sect.~\ref{sec: major recent developments and applications}, we summarise the progress made in the subsequent years after the publication of the Generation III model  \cite{emsenhuberNewGenerationPlanetary2021}, highlighting some of the most significant recent developments and applications of the model.

\section{Planetary formation and evolution framework}
\label{sec: planetary formation and evolution framework}

The Bern Model of planet formation and evolution is a global low-dimensional model that considers a large number of physical processes in a self-consistently coupled way. The Bern Model originates originally from the model first presented in Alibert et al.  \cite{alibertMigrationGiantPlanet2004,alibertModelsGiantPlanet2005}, which simulates the formation of single giant planets from the protoplanetary disk until the gas disk disperses. At the core of this model lie the planetary structure equations and the viscous evolution equation of the protoplanetary disk, which are solved numerically in the one-dimensional spherical respectively axisymmetric approximation. Inspired by the pioneering work of planetary population synthesis by Ida \& Lin \cite{idaDeterministicModelPlanetary2004}, the model was then applied in a single planet population synthesis approach for the first time in Mordasini et al. \cite{mordasiniExtrasolarPlanetPopulation2009}. Subsequently, the model developed into a single-planet long-term evolution branch on one side, and a multi-planet formation model but without long-term evolution on the other side. The planetary evolution model follows the thermodynamic cooling and contraction, as well as atmospheric escape on a billion-year timescale \cite{mordasiniCharacterizationExoplanetsTheir2012,mordasiniCharacterizationExoplanetsTheir2012a,jinPlanetaryPopulationSynthesis2014}. The formation model was developed further to be capable of simulating the formation of multi-planetary systems considering N-body interactions in three dimensions, including improved versions of the gas and planetesimal disks, as well as more advanced migration prescriptions \cite{alibertTheoreticalModelsPlanetary2013, fortierPlanetFormationModels2013,dittkristImpactsPlanetMigration2014}. In the following years, as part of the theoretical modelling efforts of the NCCR PlanetS during its phases 1 and 2, the two formation and evolution branches of the model were combined into what is  known as the Generation III Bern Model in Emsenhuber et al. \cite{emsenhuberNewGenerationPlanetary2021}, linking the formation of multi-planetary systems from the protoplanetary disk to the dispersal of the gas disk with the subsequent billion year interior thermodynamical evolution, allowing to predict all major observable characteristics of (extrasolar) planets. Due to the heritage of the model, it remained computationally manageable, capable of performing large-scale population syntheses. For the first time, this allowed for the statistical comparison of synthetic multi-planetary systems, modelled from observed initial properties of protoplanetary disks, with the long-evolved systems observed today. The Bern Model has been described in detail and applied extensively in the \emph{Next Generation Planetary Population Synthesis} (NGPPS) paper series of to date {eight} papers \cite{emsenhuberNewGenerationPlanetary2021,emsenhuberNewGenerationPlanetary2021a,schleckerNewGenerationPlanetary2021,burnNewGenerationPlanetary2021,schleckerNewGenerationPlanetary2021a,mishraNewGenerationPlanetary2021a,NGPPS7,NGPPS8}, representing a milestone for state-of-the-art planetary population synthesis as a result of a large computational effort. After the first paper introducing the model \cite{emsenhuberNewGenerationPlanetary2021}, the second NGPPS II paper \cite{emsenhuberNewGenerationPlanetary2021a} characterised formation pathways of different sub-populations of planets and compared to the observed exoplanetary population. The third publication \cite{schleckerNewGenerationPlanetary2021} studied the correlation between the presence super-Earths and giant planets in the same system. In the fourth paper  \cite{burnNewGenerationPlanetary2021}, planetary system formation around low-mass stars is considered. A data-driven approach to linking formation outcomes with initial conditions is employed in NGPPS V \cite{schleckerNewGenerationPlanetary2021a}. In NGPPS VI \cite{mishraNewGenerationPlanetary2021a}, the architectures of planetary systems are analysed under the lens of the observed peas-in-a-pod trend. The paper NGPPS VII \cite{NGPPS7} compared quantitatively the statistics and demographics found observationally by the HARPS/Coralie GTO radial velocity survey \cite{mayorHARPSSearchSouthern2011} with those in the synthetic populations. It illustrates an important synergy within the NCCR PlanetS existing between the observational efforts led by Geneva Observatory and the theoretical ones at the University of Bern. Finally, in NGPPS VIII \cite{NGPPS8} the impact of host star metallicity on a number of statistical properties (like the morphology of the radius valley) was studied and compared with observations, mainly from the Kepler transit survey.  

\subsection{Model overview}
\label{sec: model description}
The following brief overview is based on the Generation III model version presented in NGPPS I \cite{emsenhuberNewGenerationPlanetary2021}, highlighting the wealth of physical processes which are considered in the Bern Model, showcasing the significant progress that was made in the last decade. Naturally, most of the descriptions are a summary of what can be found in \cite{emsenhuberNewGenerationPlanetary2021} in  more detail.

\subsubruninhead{Star and protoplanetary disk}

The disk part of the Bern Model is based on solving the 1D radially symmetric viscous diffusion equation \cite{lustEntwicklungUmZentralkorper1952,lynden-bellEvolutionViscousDiscs1974}
\begin{equation}
    \dot{\Sigma} = \frac{3}{r}\frac{\partial}{\partial r}\left[ r^{1/2}\frac{\partial}{\partial r}\left(\nu\Sigma r^{1/2}\right)\right] - \dot{\Sigma}_{\mathrm{PEW}, \mathrm{int}} - \dot{\Sigma}_{\mathrm{PEW} ,\mathrm{ext}}- \dot{\Sigma}_\mathrm{acc} \label{eq: bm viscous disk evolution}
\end{equation}
to compute the time evolution of the protoplanetary gas disk surface density $\Sigma$ at an orbital distance $r$, considering the parametrised viscosity $\nu=\alpha c_s H$ \cite{shakuraBlackHolesBinary1973}, where $c_s$ is the isothermal sound speed and $H=c_s/\omegak$ is the vertical disk scale-height, $\omegak$ being the Keplerian frequency. {The different term on the right hand side represent, respectively, the viscous diffusion term, the sink terms due to internal and external photo-evaporation ($\dot{\Sigma}_{\mathrm{PEW}, \mathrm{int}}$ and $\dot{\Sigma}_{\mathrm{PEW} ,\mathrm{ext}}$) and the removal of gas $\dot{\Sigma}_\mathrm{acc}$ by gas-accreting planets. The radial dependance of the photo-evaporation terms is taken from \cite{ClarkeUVSwitch2001} and \cite{MatsuyamaViscousDiffPhotoevap2003}, whereas the gas-accretion term (last one in the Equation) is directly given by the gas accretion from forming planets}. The approach of Nakamoto \& Nakagawa \cite{nakamotoFormationEarlyEvolution1994} is followed for the vertical disk structure linking the midplane temperature to the disk surface temperature, considering viscous heating, the irradiation of the flared disk surface \cite{huesoEvolutionProtoplanetaryDisks2005}, as well as the direct heating of the midplane by the host star.\\

The recently manifesting paradigm shift towards MHD wind-driven disk evolution, as opposed to the classical viscous description, was introduced to the model framework in Weder et al. \cite{wederPopulationStudyMHD2023}. In Sect.~\ref{sec: mhd wind-driven disks}, we give a more detailed description of these developments and their impact on the disk evolution.

\subsubruninhead{Planetary structure and gas accretion}

The  classical 1D radially symmetric planetary internal structure equations \cite{bodenheimerCalculationsAccretionEvolution1986d} of mass conservation, hydrostatic equilibrium, and energy transport are solved for every planet at each timestep. Comparing the adiabatic and radiative gradients allows considering both convective and radiative zones within the planetary envelope. The radiative gradient crucially depends on the envelope opacity and the luminosity which is mainly due to the accretion of solids, gas and cooling and contraction of the envelope. Contributions of radioactive decay, bloating of close-in planets, and deuterium burning for the most massive planets are also considered. By solving the internal structure equations, the crucial feedback of the core accretion rate via the luminosity on the accretion rate of gas is captured, which is not self-consistently accounted for in analytical descriptions of the gas accretion rate.

The maximum gas accretion rate, past which planets accrete a large amount of gas, is assumed to be reached when the  gas accretion rate obtained from the structure equations (which is governed by the envelope's KH-contraction) exceeds the disk-limited gas accretion rate, allowing the formation of giant gas-dominated planets. The planet changes from the attached into the detached phase \cite{Bodenheimer2000,Mordasini2012charactI}.

\subsubruninhead{Planetesimal disk and accretion}

The planetesimals are described statistically on an axisymmetric 1D radial grid as  function of orbital distance from the star and time  \cite{fortierPlanetFormationModels2013}, where the planetesimals disk is modelled by a surface density $\Sigma_\text{plan}$ with a dynamical state characterised by the Keplerian orbital elements $e$ (eccentricity) and $i$ (inclination). The dynamical state of the planetesimals is evolved considering the gas drag and the stirring due to protoplanets and planetesimals alike. The planetesimal accretion rate can be expressed \cite{chambersSemianalyticModelOligarchic2006}
\begin{equation}
    \mdotplan = \omegak \bar{\Sigma}_\text{plan}\rhill^2p_\text{coll},
\end{equation}
where $\bar{\Sigma}_\text{plan}$ is the mean planetesimal surface density in the feeding zone around the planet \cite{fortierPlanetFormationModels2013} and $\rhill$ is the Hill radius. Following Inaba et al. \cite{inabaHighAccuracyStatisticalSimulation2001}, the collision probability $p_\text{coll}$ contains the kinematic details such as the gravitational focussing of the collision cross section \cite{safronovEvolutionProtoplanetaryCloud1972}. Apart from the mass of the planet and the dynamical state of the local planetesimals, the collision probability also depends on the so-called capture radius of the planet, representing the effective radius at which planetesimals can be accreted given their dynamical state \cite{podolakInteractionsPlanetesimalsProtoplanetary1988,inabaEnhancedCollisionalGrowth2003,Mordasini2006tafp}. This is important for embedded planets that have an envelope, where the capture radius can be significantly enhanced. The size of planetesimals is crucially important, as it strongly impacts the collision probability. Small planetesimals are more strongly affected by gas drag which leads to lower relative velocities and larger capture radii, resulting in larger accretion rates.

In the NGPPS series \cite{emsenhuberNewGenerationPlanetary2021}, the planetesimal radius is a fixed parameter and inward drift of planetesimals due to gas drag is neglected. A more realistic treatment involves a size distribution, the fragmentation into smaller planetesimals, and the gas drag induced drift. These crucial developments to the planetesimal disk description were introduced in Kaufmann \& Alibert \cite{kaufmannInfluencePlanetesimalFragmentation2023} and shown to fundamentally impact the accretion of planetesimals. We refer to Sect.~\ref{sec: planetesimal fragmentation} for a more detailed account of these developments.

\subsubruninhead{Orbital migration}

The exchange of angular momentum between a planet and the disk leads to orbital migration \cite{goldreichExcitationDensityWaves1979,linTidalInteractionProtoplanets1986}. Planets that are embedded in the gas disk migrate in the type I regime, which is driven by the torque exerted by spiral density waves launched at the Lindblad resonances, as well as by the co-rotation torque due to density asymmetries in the co-orbital region. The type I torque \cite{colemanFormationPlanetarySystems2014} including the attenuating effects of the planetary eccentricity and inclination \cite{bitschOrbitalEvolutionEccentric2010,cresswellThreedimensionalSimulationsMultiple2008} is considered, as well as the saturation of the co-rotation torque which occurs when the co-orbital asymmetries are no longer stable. When a massive planet opens a gap tidally, the acting torques are reduced due to the depleted surface density in the gap, leaving the planet to migrate in the so called type II regime \cite{cridaWidthShapeGaps2006}. Here, planets typically migrate slower, allowing for giant planets to remain outside the inner disk region.

Orbital migration has received a lot of attention in the last decade as one of the key processes in planet formation, shaping the emerging planetary populations. We showcase this in more detail in Sect.~\ref{sec: Planetary migration on a population level}.



\subsubruninhead{N-body and collisions}

The formation of multi-planetary systems, as opposed to a single planet per disk, represents a significant increase in computational complexity and required resources. Global models thus initially employed the one-embryo-per-disk approximation \cite{idaDeterministicModelPlanetary2004,mordasiniExtrasolarPlanetPopulation2009}. The mutual gravitational interactions between planets affect the planetary system architecture due to changes in the orbital migration patterns, ejections of planets in close encounters, competition for the accretion of gas and solid building blocks, and orbital resonances. Mutual collisions of planets, so-called giant impacts, fundamentally reshape the emerging planetary systems and provide a way for planetary cores to grow that is not possible for isolated planets. The computational complexity of the N-body problem is the most constraining aspect of global planet formation modelling. Retaining the ability to perform large-scale population synthesis for multi-planetary systems despite the increasingly broadening scope of the Bern Model remains a focus of current and future developments, especially regarding faster approaches to model the N-body interactions \cite{Grimm2022,Kimura2025}.

The gravitational interactions between planets are modelled in 3D where the contributions from disk-driven migration and gas drag are considered as additional forces. The impact of N-body interactions for the formation of planetary systems is discussed in more detail in NGPPS I and II \cite{emsenhuberNewGenerationPlanetary2021, emsenhuberNewGenerationPlanetary2021a}.

\subsubruninhead{Planet evolution}

After a pre-defined formation epoch (set to 20 Myr in the original NGPPS series, and extended to 100 Myr in the ``longshot'' simulations \cite{emsenhuberPopulationSynthesisDisks2023}), only the interior thermodynamic evolution of  single planets is followed in the subsequent billions of years of evolution by solving the internal structure equations for every planet \cite{mordasiniCharacterizationExoplanetsTheir2012a}, considering the photo-evaporative atmospheric escape due to X-ray and EUV irradiation of the star \cite{jinPlanetaryPopulationSynthesis2014}. This is crucially important in order to link the synthetic planetary systems with observations of exoplanetary systems around billions of years old stars, in particular regarding the radius and luminosity of the planets.



\subsubruninhead{Initial conditions}

The Bern Model requires a number of parameters to initialise a simulation. Some of the most important, usually fixed, parameters are the stellar mass, the viscosity parameter $\alpha$, and the planetesimal size. The planetary embryos, which are the N-body objects for which  the planetary structure equations are solved and their gas and solid accretion is calculated, are typically inserted randomly at a fixed mass $M=10^{-2}~\mearth$ at the beginning of every simulation. The parameters which are not fixed but sampled from statistical distributions for the purpose of population synthesis are the initial gas disk mass, the inner disk radius, the external photo-evaporation parameter, and the initial bulk dust-to-gas ratio. The distributions are informed by dust continuum disk mass measurements, stellar rotation periods, protoplanetary disk lifetimes, and stellar metallicities, respectively \cite{emsenhuberNewGenerationPlanetary2021a,testiProtoplanetaryDiskPopulation2022,andrewsObservationsProtoplanetaryDisk2020}.



\subsection{Planetary migration on a population level}
\label{sec: Planetary migration on a population level}

Although large-scale orbital migration remains difficult to test observationally, the presence of mean-motion resonances in multi-planetary systems is strong evidence for some degree of planetary migration \cite{weissArchitecturesCompactMultiplanet2022}. Supported by this and backed by the theoretical predictions linked to angular momentum conservation, orbital migration has become a key part of planet formation models in the last decade. We have reached a point where there are many different prescriptions to compute type I \cite{massetSaturatedTorqueFormula2010,paardekooperTorqueFormulaNonisothermal2010,paardekooperTorqueFormulaNonisothermal2011,jimenezImprovedTorqueFormula2017,guileraThermalTorqueEffects2019} and type II \cite{linTidalInteractionProtoplanets1986,syerSatellitesDiscsRegulating1995,armitageMassivePlanetMigration2007,duffellMigrationGapopeningPlanets2014,durmannMigrationMassivePlanets2015,durmannAccretionMigratingGiant2017,kanagawaRadialMigrationGapopening2018} migration rates. However, the way in which migration influences the overall planet formation process and the emerging planetary systems is not straightforward. In particular, since the migration rate strongly depends on the planet’s mass, understanding the coupling between accretion and migration is a delicate subject that requires consistent modelling of all related processes. As we illustrate in the following, the emerging planetary populations are profoundly affected by orbital migration which represents one of the most important findings in planet formation theory of the past decade.\\

\begin{figure}[t]
\centering
\includegraphics[width=\linewidth]{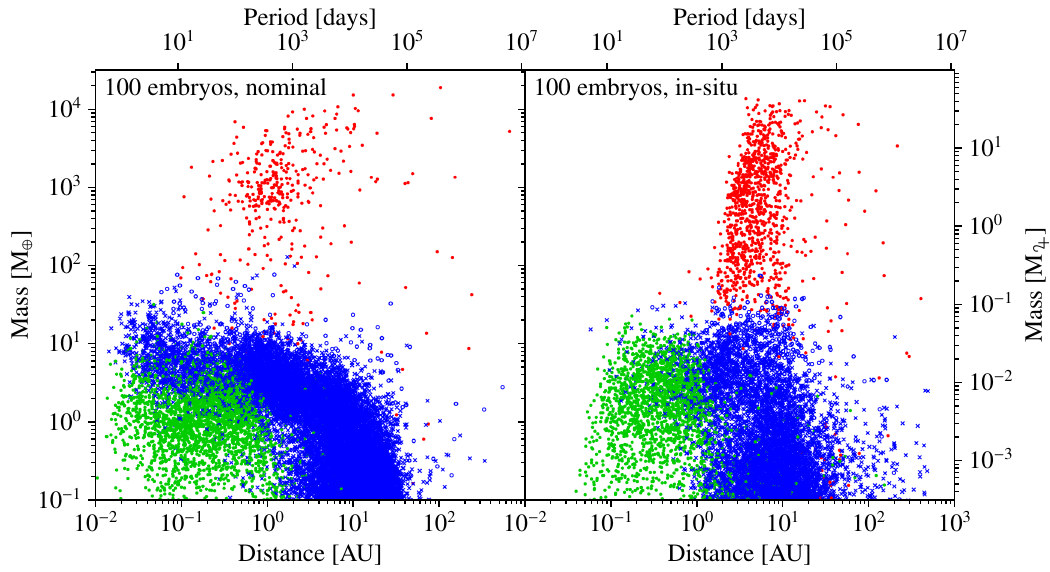}
\caption{Mass-distance diagrams for populations with initially 100 embryos ($10^{-2}~\mearth$) per disk, considering orbital migration (left) and the in-situ case (right). The colours and shapes of the symbols show the bulk composition: red points are giant planets with $\menv/\mcore>1$, blue symbols are planets that have accreted some volatile material (ices) outside of the iceline(s), and green symbols are planets that have only accreted refractory solids. Open green and blue circles have 0.1$\leq\menv/\mcore\leq1$ while filled green points and blue crosses have $\menv/\mcore\leq0.1$. Black crosses show the Solar system planets.}
\label{fig: alexandre_comparison}
\end{figure}

The Bern Model offers the possibility to examine the population-level effects of planetary migration, as well as the fundamental interactions between the disk evolution, the accretion processes, the disk-driven migration, as well as the n-body interactions between the planets. In order to characterise the effects of migration, it is helpful to compare the emerging populations to the so-called in-situ populations where the orbital migration is neglected completely. In Fig.~\ref{fig: alexandre_comparison}, this comparison is provided for populations with 100 initial planetary embryos. Without type I and II migration (in-situ, right), the final locations of the planets are clearly different to the nominal migrating case (nominal, left). This is true for rocky planets (green), planets with a gaseous envelope (blue), as well as giant planets (red).\\
Without migration, giant planets are only found close to the ice lines and beyond where planetesimal accretion is most efficient in forming massive cores. We can also note that there are giant planets and many otherwise massive bodies at locations close to the outer limit of the domain we place embryos initially (40 au), while when migration is included we find fewer giant planets beyond $\sim10$ au. Thus, the well-known problem of forming giant planets at large separations is not explained by the time needed to build a core compared to the lifetime of the gas disk as often assumed, but rather the ratio between accretion and migration timescales \cite{emsenhuberPlanetaryPopulationSynthesis2023}. Similarly, looking at planets between about 20 to 100 Earth masses when migration is neglected, there is an almost continuous population of planets throughout this mass range. This suggests that although these planets are accreting large quantities of gas, the accretion timescales are not much shorter than the lifetime of the gas disk, and the observed (sub-)Saturn desert is not well-explained by in-situ formation. We see that strong migration can impede the formation of giant planets since, with migration most efficient in the $10-50~\mearth$ range (just before passing into type II migration), planets must grow quickly to avoid being transported to the inner edge of the gas disk. This is also a well-documented issue in pebble-based planet formation scenarios, where planets in this mass range are readily formed \cite{bruggerPebblesPlanetesimalsOutcomes2020,kesslerInterplayPebblePlanetesimal2023a}. This leaves the $20-100~\mearth$ range underpopulated, while producing the large population of planets in the inner disk region which originate from the outer disk (blue). It is here, where the compositional analysis of (sub-)Neptunes on short orbits may (either via their bulk composition as expressed in the mass-radius relation of via the atmospheric composition obtained from transit spectroscopy) provide a way to test large-scale inward migration observationally. The emerging picture that the radius valley \cite{fultonCaliforniaKeplerSurveyIII2017} is caused by the different formation pathways of smaller dry in-situ super-Earth versus larger water-rich ex-situ  sub-Neptunes that have migrated in from beyond the iceline would give support to the large-scale migration hypothesis \cite{venturiniNatureRadiusValley2020,burnRadiusValleyMigrated2024}.\\
Considering which protoplanetary disks form giant planets reveals that since embryos have to start further away when Type I migration is stronger, the systems must include a higher mass of solids in the first place. Dynamical timescales at large separations are longer, so the collision probabilities between embryos and planetesimals are smaller overall. To compensate for this, there must be a higher amount of planetesimals available, skewing the giant planet-forming disks further towards higher metallicity disks when migration is considered. Given the same initial metallicity distribution used for the populations shown in Fig.~\ref{fig: alexandre_comparison}, this is linked to the lower amount of giant planets formed in the nominal setup.\\

In summary, orbital migration does not only simply shift the emerging planetary population compared to the in-situ formation of planets. The formation history, the planetary mass function, the metallicity dependence, etc., also change due to the interactions between the disk evolution, the core accretion processes, and the n-body evolution.



\subsection{Success and limitations of Bern population synthesis}

{The Bern population synthesis, in particular the NGPPS series, were extensively compared to different types of observations, mainly from Kepler and from radial-velocity surveys. These comparisons showed a fair agreement between the NGPPS populations and Kepler results, in particular regarding the period-radius distribution of planets \cite{mulders2019} as well as the `peas-in-a-pod' pattern \cite{Mishra_etal_2021}. Comparison with the HARPS and CORALIE surveys also demonstrated a good agreement, like the presence of two groups of planets in the mass-distance diagram, namely the close-in sub-Neptunes and distant giants, the existence of a bimodal mass function with a less populated 'desert' \cite{NGPPS7,NGPPS8}. In addition, the mean multiplicity of detectable planet in synthetic systems is about 1.6, similar to the value given by observations. This is relevant, as multiplicity is a first proxy for system architecture.}

\begin{table}[ht]
\centering
\caption{Some of the successes of the NGPPS Bern Model simulations}
\label{successes}
\begin{tabularx}{\textwidth}{lXX}
\hline
Feature & Model Prediction & Observational Confirmation \\
\hline
Low-mass planet prevalence & 
High abundance of small planets \cite{Alibert_etal_2013,NGPPS7} & 
Confirmed by high-precision radial velocity surveys and Kepler \\
Radius distribution & 
Pile-up around 1 Rj \cite{Mordasini_etal_2012b} & 
Observed in cleaned Kepler data \\
Evaporation valley & 
Depleted region in distance-radius plane 
\cite{jinPlanetaryPopulationSynthesis2014} & 
Confirmed by multiple observations \cite{fultonCaliforniaKeplerSurveyIII2017}\\
Multi-planet architectures & 
Co-planar sub-Neptune systems at short periods \cite{Mishra_etal_2021} & 
Matches Kepler super-Earth population \\
Period ratio distribution & 
100-seed model NGPPS prediction \cite{dichangchenperiodratios2024,NGPPS8} & 
Closely matches Kepler distribution except for $\sim$2 times too many planets near MMR \cite{weissArchitecturesCompactMultiplanet2022} \\
\hline
\end{tabularx}
\end{table}

{Some notable differences where however identified. In the comparison with RV surveys \cite{NGPPS7}, for instance, too many planets were predicted by the models, which is though to arise because of the efficiency of solid accretion or the initial conditions. The planetary `desert' was found to be too deep by $\sim 60 \%$, which is likely arising from a too high gas accretion rate in the corresponding mass range. Planet eccentricities were found to be too low in the synthetic population, which could arise from the damping while being in the gas disk and/or mechanisms increasing the eccentricities including dynamical excitation  over long timescales. The comparison with both RV and Kepler data showed that simulations predict too many planets in resonances \cite{dichangchenperiodratios2024} and planets at too short orbital periods (innermost planets of multi-planet systems are a factor four too close to their star \cite{mulders2019}, which is thought to arise because of resonance chain breaking during the dispersion of the gas disk and the halting of migration because of substructures in the disk, respectively.}

{These limitations, as well as theoretical and observational progress, have motivated the developments that are outlined in the next section.}

\section{Major recent developments and applications}
\label{sec: major recent developments and applications}

The Bern Model has since been developed further to include new physics and the model, parts of it, and modified versions, have been applied in a number of diverse studies in the subsequent years after the publishing of the NGPPS paper series. \cite{bruggerPebblesPlanetesimalsOutcomes2020} compared planetesimal and pebble-based planet formation scenarios. \cite{mollousPotentialLongtermHabitable2022} investigated the long-term habitability of planets with primordial H-He envelopes. \cite{voelkelMultipleGenerationsPlanetary2022a} studied the dynamic formation of planetary embryos and the emerging multiple generations of planets in the same system. The Bern Model was used to investigate concurrent pebble and planetesimal accretion in \cite{kesslerGiantFormationPebbles2022}. \cite{davoultEarthlikePlanetsHosting2024} analysed the architectures of planetary systems emerging from the Bern Model with a focus on systems that are hosting Earth-like planets. \cite{burnRadiusValleyMigrated2024} studied the radius valley between super-Earths and sub-Neptunes using an improved treatment of the planetary composition based on new equations of state of water \cite{haldemannAQUACollectionH2O2020}. \cite{shibaikeConstraintsPDS702024} implemented a simple model of a circumplanetary disk to constrain the planetary properties from the dust emission measurements in the PDS70 system, where two planets have been successfully identified while still undergoing formation \cite{gaiacollaborationGaiaDataRelease2018,mullerOrbitalAtmosphericCharacterization2018}. Using synthetic populations from the Bern Model, \cite{eggerUnveilingInternalStructure2024} constrain the formation history and disk conditions of an exoplanet system observed with the CHEOPS telescope. \cite{kaufmannStreamingInstabilityOnset2025} investigate the early planet formation phase in a ring of planetesimals, formed from the streaming instability. In order to calculate observable disk properties, an improved treatment of dust dynamics and the calculation of continuum dust emissions was implemented in \cite{burnPopulationSynthesisDisks2022a}. In \cite{emsenhuberPopulationSynthesisDisks2023}, the best-fitting initial disk conditions have been evaluated in a parameter study, focussing on observed disk masses and lifetimes. This demonstrates the capability of the Bern Model to perform disk population syntheses as well, which enables the detailed and coupled analysis of the disk evolution and the formation of planets on the level of large populations.\\

In the following, we highlight some of the major additions to the model: the wind-driven disk evolution model and the updated external photo-evaporation description \cite{wederPopulationStudyMHD2023}, the fragmentation and drift of planetesimals \cite{kaufmannInfluencePlanetesimalFragmentation2023}, the inclusion of the evolution of dust, planetesimal formation, and pebble accretion \cite{voelkelEffectPebbleFluxregulated2020}, as well as the presence of compositional gradients in planetary envelopes in the evolution phase \cite{polmanConvectiveMixingDistant2024}. Finally, we show how the study of the emerging planetary system architectures enhances our understanding of planet formation processes and further enable predictive modelling.


\subsection{MHD wind-driven disks}
\label{sec: mhd wind-driven disks}

The evolution of the protoplanetary disk is of fundamental importance as it sets the stage of planet formation. Until recently, planet formation has been studied almost exclusively in disks that evolve due to turbulent viscosity \cite{drazkowskaPlanetFormationTheory2022a}. Relatively high levels of turbulence are needed to obtain observed accretion rates \cite{hartmannAccretionEvolutionTauri1998}. A disk threaded by a magnetic field is susceptible to magneto-rotational instability (MRI) \cite{balbusPowerfulLocalShear1991,balbusInstabilityTurbulenceEnhanced1997} which induces turbulence. While in the inner disk where ($T>1000\,\mathrm{K}$) and in the outer disk where ($\Sigma \lesssim 15\,\mathrm{gcm^{-2}}$) the conditions are met for MRI to be sustained, non-ideal MHD effects such as Ohmic and ambipolar diffusion suppress MRI in large parts of the disk from $1$ to several tens $\mathrm{AU}$ \cite{lesurHydroMagnetohydroDustGas2022}. However, an emerging MHD wind is still able to remove angular momentum from the disk, driving accretion \cite{blandfordHydromagneticFlowsAccretion1982,koniglEffectsLargeScaleMagnetic2010}. This leads to disks evolving due to a laminar accretion flow either in a surface layer or through the midplane \cite{perez-beckerSurfaceLayerAccretion2011,perez-beckerSURFACELAYERACCRETION2011a,baiWINDDRIVENACCRETIONPROTOPLANETARY2013,gresselGlobalHydromagneticSimulations2020}.

In a scenario, where disk evolution is primarily driven through angular momentum being removed by an emerging MHD wind, contrary to the viscous scenario where angular momentum is redistributed outwards, the evolution equation \eqref{eq: bm viscous disk evolution} is extended by an advection term and a sink term from the emerging MHD wind \cite{suzukiEvolutionProtoplanetaryDiscs2016}
\begin{equation}
    \begin{aligned}
        \dot{\Sigma}= & \frac{1}{r} \frac{\partial}{\partial r}\left[\frac{3}{r \Omega} \frac{\partial}{\partial r}\left(r^2 \Sigma \overline{\alpha_{r \phi}} c_{\mathrm{s}}^2\right)\right]+\frac{1}{r} \frac{\partial}{\partial r}\left[\frac{2}{\Omega} r \overline{\alpha_{\phi z}}\left(\rho c_{\mathrm{s}}^2\right)_{\text {mid }}\right] \\
        & -\dot{\Sigma}_{\mathrm{MDW}}-\dot{\Sigma}_{\mathrm{PEW}, \mathrm{int}}-\dot{\Sigma}_{\mathrm{PEW}, \mathrm{ext}} - \dot{\Sigma}_\mathrm{acc}.
    \end{aligned}
\end{equation}
Here, accretion is driven by the magnetic stress $\overline{\alpha_{\phi z}}$ and $\overline{\alpha_{r\phi}}$ corresponds the background effective turbulent viscosity which is generally low. The mass-loss-rate from the MHD wind is connected to the stellar accretion rate through the amount of angular momentum the wind can remove \cite{pascucciRoleDiskWinds2023}. Mass-loss profiles from the emerging wind can be obtained by assuming an accretion powered wind, meaning that the wind is calculated self consistently by calculating the liberated gravitational energy from accretion and assuming part of the liberated energy goes into launching the wind while the rest goes into heating the disk \cite{suzukiEvolutionProtoplanetaryDiscs2016}. In this scenario, a strong wind corresponds to a wind that carries less angular momentum and vice versa.

This leads to many differences in the evolution of the disk as opposed to the viscous scenario. A strong wind will lead to overall flatter surface density profiles and depending on the evolution of $\overline{\alpha_\phi z}$ can even lead to inverted surface density profiles in the inner disk. Further, internal photo-evaporation through EUV and/or X-ray photo-evaporation can be suppressed by shielding of the emerging wind, leading to an early phase where the evolution is dominated by the MHD wind and a late phase where the disk is dissipated by internal photo-evaporation \cite{kunitomoDispersalProtoplanetaryDiscs2020,pascucciRoleDiskWinds2023,wederPopulationStudyMHD2023}. Furthermore, the temperature profiles can differ substantially from the viscous case, as there is less viscous heating in the midplane \cite{moriTemperatureStructureInner2019}.

The new disk evolution paradigm further has many implications for planet formation, in particular planetary migration. For low-mass planets, that are in the type I migration regime, the low turbulent viscosity leads to an early saturation of the corotation torque that would lead to faster inward migration. It gives also rise to the dynamical corotation torque, that can have different properties depending on the vortensity profile and the flow configuration in the disk (i.e. layered accretion or laminar flow through the midplane) \cite{paardekooperTorqueFormulaNonisothermal2011,paardekooperDynamicalCorotationTorques2014,mcnallyLowMassPlanet2017,mcnallyLowmassPlanetMigration2018,mcnallyLowmassPlanetMigration2020}. A flat or reverted surface density profile can slow down or even revert inward migration \cite{ogiharaSuppressionTypeMigration2015}. Due to the lack of turbulent viscosity, gap opening can happen earlier at lower masses and the gaps are generally deeper \cite{paardekooperPlanetDiskInteractions2022,aoyamaThreedimensionalGlobalSimulations2023a}. The interaction of a gap opening planet with the accretion flow and the magnetic field is still subject to current research as no clear picture has emerged yet \cite{kimmigEffectWinddrivenAccretion2020,legaMigrationJupiterMass2022,nelsonGasAccretionJupiter2023,aoyamaThreedimensionalGlobalSimulations2023a,wafflard-fernandezPlanetdiskwindInteractionMagnetized2023,wafflard-fernandezGoneWindOutward2025}.

\subsection{Planetesimal fragmentation}
\label{sec: planetesimal fragmentation}

Since the accretion rate of planetesimals is highly sensitive to their size, as discussed in chapter \ref{sec: model description}, the accurate modelling of the evolution of the planetesimals is vital to understand planet formation. Many insights from the investigation of planetesimal formation \cite{schaferInitialMassFunction2017,abodMassSizeDistribution2019} and observational constraints from the solar system \cite{morbidelliAsteroidsWereBorn2009} point to initial planetesimal sizes of $\sim 100$ km. This poses a challenge for planetesimal-based formation models as they struggle to explain the formation of giant planets within the lifetime of the protoplanetary disk due to long accretion timescales.

To improve on the previously used description of the planetesimals with a fixed size of $300$ m, we introduce a fragmentation model based on the works of \cite{ormelUnderstandingHowPlanets2012} that accounts for the change in the typical size of planetesimals during the formation due to their mutual collisions. Typically, in the later stages of planet formation, these collisions are energetic enough to lead to fragmenting collisions, reducing the average size of planetesimals. These smaller collisional remnants, called fragments ($\Sigma_\text{f}$), are produced at a rate that can be described by 
\begin{equation}
        \label{eq:sigmadot_frag}
        \dot{\Sigma}_\text{f} = \frac{q_\text{plan}}{1+q_\text{plan}} \frac{\Sigma_\text{plan}}{\text{T}_\text{plan}} + \frac{q_\text{plan,f} \Sigma_\text{plan}}{\text{T}_\text{plan,f}}
\end{equation}
where $q_{i,k}$ is the ratio of the size-dependent collisional energy and specific material strength \cite{benzCatastrophicDisruptionsRevisited1999} and $T_{i,k}$ are the collisional timescales between the relevant components. From the outcome of the collisional cascade, we expect the typical size of these fragments to be ($\sim 100$ m) but it strongly depends on the assumed material properties and we consider multiple treatments for their typical size \cite{kaufmannInfluencePlanetesimalFragmentation2023}. Treating the full collisional evolution has been explored in multiple works as well \cite{guileraPlanetesimalFragmentationGiant2014,kaufmannStreamingInstabilityOnset2025}, but proved to be too computationally expensive for population synthesis studies. Similar to pebbles, these smaller collisional fragments are more closely coupled to the gas, resulting in significant radial drift which leads to a redistribution of the planetesimals in the disk and depleting the feeding zone of the forming planets.\\

We performed a population synthesis study, accounting for these additional processes affecting the planetesimals to investigate how fragmentation affects the planet formation process \cite{kaufmannInfluencePlanetesimalFragmentation2023}. Our main findings are that for the nominal model with fragments of a fixed size ($100$ m), fragmentation promotes the growth of planetary embryos, especially beyond the ice line. However, for bigger initial planetesimals ($100$ km), planetesimal-based accretion models still fail to explain the formation of giant planets across a large space of initial conditions. Additionally, if the collisionally produced fragments are small, fragmentation can even lead to reduced growth of planetary embryos as the feeding zones of the forming planets get depleted due to the radial transport of the fragments. In summary, we found that the consideration of these additional effects is not sufficient to resolve the shortcomings of planetesimal-based formation models, i.e. the struggle to explain the formation of cold giant planets.

\subsection{Dust evolution and pebble accretion}
\label{sec: dust evolution and pebble accretion}

The evolution of dust grains into so-called pebbles, has received a lot of attention in the past 15 years, ever since Ormel \& Klahr \cite{ormelEffectGasDrag2010} demonstrated the potentially dominant role the direct accretion of pebbles by growing planets could play. Pebbles are defined by their decoupled motion from the gas dynamics which is characterised by the Stokes number $\st=t_\text{stop}\omegak$ relating the timescale on which particles adapt to the gas motion $t_\text{stop}$ with the timescale on which gas perturbations occur ($\sim1/\omegak$). While micrometre dust grains have tiny Stokes numbers and follow the gas motion in a well-coupled manner, pebbles are characterised by larger values of $\st$, approaching unity, and are increasingly decoupled from the gas motion. The significantly enhanced collisional cross sections due to the strong gas drag on particles in the pebble size range, paired with a steady supply of inwards drifting pebbles from the outside regions of the disk, result in short planet formation timescales. This has been suspected to be promising for the formation of giant planets, especially at larger separations from the star \cite{lambrechtsFormingCoresGiant2014}.

In order to model the evolution of dust into pebbles, we follow the approach of Birnstiel and collaborators \cite{birnstielSimpleModelEvolution2012a}. This model was first implemented and applied in the Bern Model framework in Voelkel et al. \cite{voelkelEffectPebbleFluxregulated2020}. Instead of considering a size distribution, the aptly named \emph{two-population} model (subscript "tp") solves the problem in a mass-averaged approach, considering only two populations of solids: the small dust grains and the larger pebbles. This is a valid approach since the dust mass is dominated by largest particles \cite{birnstielGasDustEvolution2010}. The advection-diffusion dust evolution equation is transformed into a single evolution equation for the combined two-population surface density $\Sigma_\text{tp}$
\begin{equation}
    \dot{\Sigma}_{\text{tp}} + \frac{1}{r} \frac{\partial}{\partial r} \left[ r \left( \Sigma_{\text{tp}} \bar{v} - D_g \sigmag \frac{\partial}{\partial r} \left( \frac{\Sigma_{\text{tp}}}{\sigmag}\right) \right) \right] = 0, \label{eq: bm twopop evolution}
\end{equation}
where $\bar{v}$ is the mass-averaged radial drift speed of the small and the large populations and the mass-averaged diffusivity is assumed to be equal to the gas diffusivity $D_g$. The combined surface density $\Sigma_\text{tp}$ separates into the surface densities of the two underlying populations $\Sigma_{\text{peb}}$ and $\Sigma_{\text{grains}}$, according to a factor that is calibrated with full dust evolution simulations \cite{birnstielGasDustEvolution2010}.

There exist many valid prescriptions for the pebble accretion rate which agree within orders of unity \cite{ormelEffectGasDrag2010,ormelUnderstandingHowPlanets2012,lambrechtsRapidGrowthGasgiant2012,lambrechtsFormingCoresGiant2014,guillotFilteringProcessingDust2014, idaRadialDependencePebble2016, johansenFormingPlanetsPebble2017}. Apart from the planetary mass, pebble accretion depends fundamentally on the supply of pebbles. This is strongly linked to the size of pebbles which, together with the radial gas disk pressure structure, dictates the inward drift speed of pebbles. Pebble accretion rates can become especially high, when a planet is able to accrete pebbles from the full vertical extent of the disk of pebbles, revealing an additional dependence on the (vertical) disk turbulence. Pebble accretion stops when a planet grows beyond the so-called pebble isolation mass \cite{lambrechtsFormingCoresGiant2014,bitschGrowthPlanetsPebble2015}, where the planet accelerates the gas to (super-)Keplerian velocities, halting the inward drift of pebbles.\\

The implications of pebble accretion as opposed to planetesimal accretion are manifold. In Brugger et al. \cite{bruggerPebblesPlanetesimalsOutcomes2020}, planetesimal and pebble-based planet formation scenarios were compared \textbf{using the exact same initial conditions (disk surface density profile, starting location of embryos - assumed to be located at random following a uniform-in-log distribution, solid-to-gas ratio, etc.)}, observing a strong impact of the different accretion rates on the migration of planets and, therefore, on the planet formation outcome. Pebble accretion promotes the formation of close-in super-Earths, whereas the accretion of small planetesimals promotes the formation of giant planets. This behaviour is corroborated by the study of concurrent pebble and planetesimal accretion in Kessler \& Alibert \cite{kesslerInterplayPebblePlanetesimal2023a}, where the interplay between orbital migration and the accretion luminosity prevents the formation of giant planets altogether. Coupled with a planetesimal formation model  \cite{lenzPlanetesimalPopulationSynthesis2019}, we can now simulate the formation of planets all the way from initially micrometer sized dust grains self-consistently. In \cite{voelkelEffectPebbleFluxregulated2020,voelkelMultipleGenerationsPlanetary2022a}, this approach has revealed the possibility of multiple generations of planets being formed, shaped by different accretion mechanisms and migration patterns.

\subsection{Disk structure}
\label{sec: disk structure}

The ubiquity of substructures in protoplanetary disks as revealed by ALMA \cite{huangDiskSubstructuresHigh2018} has altered our picture of the typical environment planets form in. Even though the origin of these substructures is still debated \cite{baeStructuredDistributionsGas2022}, they certainly have a huge impact on the planet formation process. This led many recent works to investigate the formation of planets at special locations in the disk. Some works consider smooth disks with localized rings of planetesimals \cite{kaufmannStreamingInstabilityOnset2025,lorekGrowingSeedsPebble2022a,liuGrowthStreamingInstability2019,jiangEfficientPlanetFormation2022} while others model the observed the dust rings by modifying the structure of the gas disk \cite{jiangEfficientPlanetFormation2022,lauRapidFormationMassive2022} or considering material infalling on the disk \cite{zhaoPlanetesimalFormationPressure2025}. The heterogeneity of the setups and uncertainty of the mechanism causing the sub-structures has not yielded a consistent picture yet, but there are a few points consistent across different works worth highlighting. If the rings are dust traps, it leads to the rapid growth of the formed planetesimals (via the streaming instability) by pebble accretion, which is very efficient in the rings due to the increased dust surface densities and the increased encounter times \cite{jiangEfficientPlanetFormation2022,lauRapidFormationMassive2022}. This serves as a promising formation pathway to explain giant planets at large separations that have been challenging for core accretion models. In disks where the gas structure is unaltered, there remains a gap between the characteristic initial planetesimal mass and the transition mass where pebble accretion becomes relevant, making early growth from the mutual collisions of planetesimals a relevant channel of growth to reach this threshold. However, it has to be noted that this transition is strongly dependent on the pebble properties and size distribution \cite{lyraAnalyticalTheoryGrowth2023}.\\

The Bern Model has been used in multiple ways to investigate the aforementioned emerging paradigms and formation scenarios. The work of Kaufmann et al. \cite{kaufmannStreamingInstabilityOnset2025} looks at the initial growth phase in planetesimal rings in smooth disks across different stellar masses to infer the timing of core formation in this scenario. We find that in smooth disks there is a significant delay to the onset of pebble accretion due to this initial growth from planetesimal collisions. This delay is strongly dependent on radial distance and stellar mass, leading to longer formation times at large radial distances and lower stellar masses. Additionally, in a currently ongoing project, we investigate the influence of late infalling material onto the protoplanetary disk and whether it can lead to the formation of planetesimals (and subsequently planets) via the streaming instability. The preliminary results show that in order for the infalling material to trigger the formation of planetesimals, the disk needs to have a low viscosity and high amounts of infalling material. However, if we consider the additional turbulence due to the infalling material \cite{kuznetsovaAnisotropicInfallSubstructure2022} the dust fails to concentrate sufficiently to trigger planetesimal formation due to the viscous evolution of the disk. 

\subsection{Compositional gradients}
\label{sec: compositional gradients}

\begin{figure}[t]
\resizebox{\hsize}{!}{\includegraphics{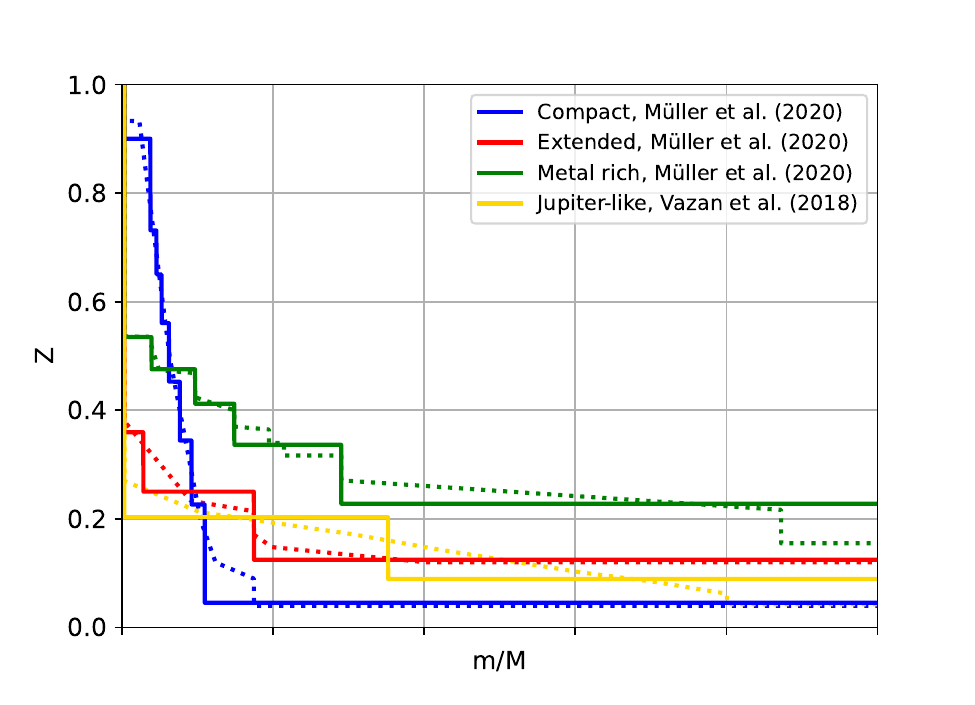}}
    \caption{Composition for the different models at the start (dotted) and end (solid) of the simulation at 4.5 Gyr. The x-axis shows the normalized mass and the y-axis the fraction of high-Z material, water in our case. The initial compositions are based on those presented in \cite{vazanJupitersEvolutionPrimordial2018} and \cite{mullerChallengeFormingFuzzy2020}.}
\label{Fig:Fig1_Polman}
\end{figure}

In recent years, measurements by the Juno mission \cite{boltonJupitersInteriorDeep2017} have shown that Jupiter's interior likely contains a dilute core \cite{wahlComparingJupiterInterior2017,debrasNewModelsJupiter2019,miguelJupitersInhomogeneousEnvelope2022,howardJupitersInteriorJuno2023} can retain complex compositional structures over their lifetime. While more complex structures have been suggested for Jupiter before \cite{stevensonFormationGiantPlanets1982,vazanCONVECTIONMIXINGGIANT2015}, the structure was often assumed to consist of a well-defined core and a convective homogeneous envelope \cite{guillotINTERIORSGIANTPLANETS2005}. These measurements combined with the launch of JWST leading to a great increase in our capability to characterise exoplanetary atmospheres and constrain their atmospheric composition \cite{guzman-mesaInformationContentJWST2020,aldersonEarlyReleaseScience2023,dyrekSO2SilicateClouds2024,bennekeJWSTRevealsCH$_4$2024} highlight the importance of correctly modelling exoplanet interiors. To understand how the atmospheric composition of an exoplanet relates to its formation history, it is essential to understand how the atmospheric composition relates to the bulk composition. 

The ability to model compositional gradients and the convective mixing that degrades them has been implemented in the evolution part of Bern Model in \cite{polmanConvectiveMixingDistant2024}. Compositional gradients have been made possible by considering the metallicity, Z, in the internal structure equations. For hydrogen and helium we use the equations of state of \cite{chabrierNewEquationState2021} and the equation of state of \cite{haldemannAQUACollectionH2O2020} for water to represent the high-Z material, in line with the rest of the Bern Model. To facilitate convective mixing according to the Ledoux criterion, the conductive opacity tables of \cite{cassisiUpdatedElectronConductionOpacities2007} have been added, whereas previously the conductive opacity was not relevant, as the envelope was assumed to be convective throughout. Similarly, the luminosity throughout the envelope has been updated from being constant to scaling with either the mass or as $\frac{dl}{dm}=-T\frac{dS}{dt}$ where $\frac{dS}{dt}$ is taken to be constant for numerical stability. The temperature gradient is computed using the mixing length theory described in the appendix of \cite{vazanCONVECTIONMIXINGGIANT2015}, considering efficient and inefficient convection, as well as semi-convection according to \cite{langerSemiconvectiveDiffusionEnergy1983,langerEvolutionMassiveStars1985}.\\

Results for on a Jupiter mass planet with an initial luminosity of 10$^3$ L$_J$ and an initial metallicity distributions based on \cite{mullerChallengeFormingFuzzy2020} and \cite{vazanJupitersEvolutionPrimordial2018} are shown in \ref{Fig:Fig1_Polman}. We find that, for these conditions, dilute cores can be retained throughout the lifetime of the planet. To see whether it is possible to distinguish a planet with a dilute core from a completely mixed planet, we compare to the same simulations with a constant metallicity throughout the envelope equal to the average metallicity of the simulations with a dilute core. We find that the effect on the radius is small, with the radius being around 5\% larger for the unmixed planet during the first 100 Myr, with this difference reducing to $\sim$1\% after the first Gyr. For the luminosity we find that the luminosity is 0.51 L$_J$ at the end of the simulation for the mixed planet and 0.63 L$_J$ for the initially unmixed planet. It is unclear whether this difference is due to differences in cooling rates or due to the unmixed planet's luminosity initially increasing due to the mixing process. Overall the effects on the radius and luminosity are small and it is unlikely that these parameters alone can be used to determine whether a planet has a dilute core, although they can provide valuable insights when combined with atmospheric characterisation.

Summarising our results we find that dilute cores can be retained for a wide variety of initial compositions, but that they also exhibit a wide variety with regards to the final extend of the dilute core and the atmospheric metallicity, highlighting the importance of understanding the composition during planet formation. Hot Jupiters can experience a decrease in convective mixing due to the decreased cooling rate of these planets, but this effect is limited. Semi-convection can alter the composition of a planet as long as the efficiency is high, but in most cases dilute cores are still retained. We find that for a 1 M$_J$ planet it is hard to retain a dilute core for luminosities higher than 3 $\times$ 10$^3$ L$_J$, although this is strongly dependent on the initial compositional structure. A more detailed explanation of the results and limitations can be found in \cite{polmanConvectiveMixingDistant2024}.

\subsection{Emerging planetary system architectures}
\label{sec: emerging planetary system architectures}

\begin{table*}[t]
\begin{center}
        \caption{Summary of the conditional probabilities of a system to host an ELP depending on its biased architecture and the properties of its IDP. Table from \cite{davoultEarthlikePlanetsHosting2024}. M$_{IDP}$, R$_{IDP}$, and P$_{IDP}$ refer to the mass, radius, and period respectively of the IDP in the system. Percentages indicate the percentage of systems hosting an ELP. N.A. corresponds to categories with too few systems to conclude. The results in bold highlight the cases with the highest probability of finding an ELP.}
        \label{tab:Davoult2024_summary}
    \begin{tabular}{l|c|c|c}
        & \textbf{G-pop (1~M$_{\odot}$)} & \textbf{earlyM-pop (0.5~M$_{\odot}$)} & \textbf{lateM-pop (0.2~M$_{\odot}$)} \\
        \hline
                    & M$_{IDP}$ $<$ 10~M$_{\oplus}$ $\Rightarrow$ 64.0\% & M$_{IDP}$ $<$ 10~M$_{\oplus}$ $\Rightarrow$ 68.5\% & \\
                    & M$_{IDP}$ $>$ 10~M$_{\oplus}$ $\Rightarrow$ 49.7\% & M$_{IDP}$ $>$ 10~M$_{\oplus}$ $\Rightarrow$ 59.2\% & \\
                  & R$_{IDP}$ $<$ 2.5~R$_{\oplus}$ $\Rightarrow$ 55.8\% & R$_{IDP}$ $<$ 2.75~R$_{\oplus}$ $\Rightarrow$ 64.1\% & \\
 \textit{Low-mass} & \textbf{R$_{IDP}$ $>$ 2.5~R$_{\oplus}$ $\Rightarrow$ 87.5\%} & \textbf{R$_{IDP}$ $>$ 2.75~R$_{\oplus}$ $\Rightarrow$ 94.6\%} & 88\% \\
                  & P$_{IDP}$ $<$ 10~days $\Rightarrow$ 38.0\% & P$_{IDP}$ $<$ 10~days $\Rightarrow$ 60.3\% & \\
                  & \textbf{P$_{IDP}$ $>$ 10~days $\Rightarrow$ 82.7\%} & P$_{IDP}$ $>$ 10~days $\Rightarrow$ 79.2\% & \\
        \hline
                  & M$_{IDP}$ $<$ 100~M$_{\oplus}$ $\Rightarrow$ 37.5\% & & \\
                    & M$_{IDP}$ $>$ 100~M$_{\oplus}$ $\Rightarrow$ 6\% &  & \\
                  & R$_{IDP}$ $<$ 10~R$_{\oplus}$ $\Rightarrow$ 34.3\% & & \\
\textit{Anti-Ordered} & R$_{IDP}$ $>$ 10~R$_{\oplus}$ $\Rightarrow$ 5.3\% & N.A. & N.A. \\
                    & P$_{IDP}$ $<$ 30~days $\Rightarrow$ 23.3\% & & \\
                    & P$_{IDP}$ $>$ 30~days $\Rightarrow$ 20.2\% & & \\
        \hline
                  & M$_{IDP}$ $<$ 10~M$_{\oplus}$ $\Rightarrow$ 30.6\% & M$_{IDP}$ $<$ 10~M$_{\oplus}$ $\Rightarrow$ 50.0\% & \\
                  & M$_{IDP}$ $>$ 10~M$_{\oplus}$ $\Rightarrow$ 7.1\% & M$_{IDP}$ $>$ 10~M$_{\oplus}$ $\Rightarrow$ 21.6\% &  \\
                & R$_{IDP}$ $<$ 6~R$_{\oplus}$ $\Rightarrow$ 27.8\% & R$_{IDP}$ $<$ 2~R$_{\oplus}$ $\Rightarrow$ 50.4\% & N.A. \\
   \textit{Ordered} & R$_{IDP}$ $>$ 6~R$_{\oplus}$ $\Rightarrow$ 2.9\% & R$_{IDP}$ $>$ 2~R$_{\oplus}$ $\Rightarrow$ 33.8\% & \\
                    & P$_{IDP}$ $<$ 30~days $\Rightarrow$ 14.1\% & P$_{IDP}$ $<$ 10~days $\Rightarrow$ 40.7\% & \\
                    & P$_{IDP}$ $>$ 30~days $\Rightarrow$ 67.5\% & P$_{IDP}$ $>$ 10~days $\Rightarrow$ 60.5\% & \\
        \hline
                & M$_{IDP}$ $<$ 10~M$_{\oplus}$ $\Rightarrow$ 31.7\% &  &  \\
                  & M$_{IDP}$ $>$ 10~M$_{\oplus}$ $\Rightarrow$ 13.1\% & & \\
                    & R$_{IDP}$ $<$ 2.5~R$_{\oplus}$ $\Rightarrow$ 27.4\% & & \\
     \textit{Mixed} & R$_{IDP}$ $>$ 2.5~R$_{\oplus}$ $\Rightarrow$ 8.33\% & N.A. & N.A. \\
                    & P$_{IDP}$ $<$ 30~days $\Rightarrow$ 13.8\% & & \\
                    & P$_{IDP}$ $>$ 30~days $\Rightarrow$ 49.1\% & & \\
        \hline
                  & \textbf{M$_{IDP}$ $<$ 100~M$_{\oplus}$ $\Rightarrow$ 94.8\%} & \textbf{M$_{IDP}$ $<$ 10~M$_{\oplus}$ $\Rightarrow$ 92.8\%} & \\
                  & M$_{IDP}$ $>$ 100~M$_{\oplus}$ $\Rightarrow$ 3.6\% & M$_{IDP}$ $>$ 10~M$_{\oplus}$ $\Rightarrow$ 90.2\% & \\
                  & \textbf{R$_{IDP}$ $<$ 8~R$_{\oplus}$ $\Rightarrow$ 95\%} & R$_{IDP}$ $<$ 2.75~R$_{\oplus}$ $\Rightarrow$ 75.1\% & \\
        \textit{n~=~1} & R$_{IDP}$ $>$ 8~R$_{\oplus}$ $\Rightarrow$ 7.6\% & \textbf{R$_{IDP}$ $>$ 2.75~R$_{\oplus}$ $\Rightarrow$ 97.3\%} & 94\% \\
                  & P$_{IDP}$ $<$ 30~days $\Rightarrow$ 37.0\% & P$_{IDP}$ $<$ 10~days $\Rightarrow$ 49.3\% &  \\
                  & \textbf{P$_{IDP}$ $>$ 30~days $\Rightarrow$ 90.6\%} & \textbf{P$_{IDP}$ $>$ 10~days $\Rightarrow$ 95.8\%} & \\

    \end{tabular}
\end{center}
\end{table*}

The study and classification of planetary system architectures are essential for a better understanding of the formation and evolution of planetary systems, including our own Solar System. The Corot \cite{auvergneCoRoTSatelliteFlight2009}, Kepler \cite{boruckiKeplerPlanetDetectionMission2010}, and now TESS \cite{rickerTransitingExoplanetSurvey2015} missions have revolutionized the study of planetary systems by detecting thousands of multi-planet systems. This increase in the detection of such systems has expanded the focus beyond just the diversity of exoplanets to also include the diversity of planetary systems. By studying exoplanet diversity, we can test and refine planetary formation models, just as studying the diversity of planetary systems allows us to improve our models of planetary system formation.

For a long time, the Solar System was our only reference for multi-planet systems, leading to the emergence of models attempting to replicate its architecture. Examples include the Grand Tack model \cite{walshLowMassMars2011,obrienWaterDeliveryGiant2014}, which reproduces the early formation and migration of the inner Solar System planets, and the Nice model \cite{gomesOriginCataclysmicLate2005,tsiganisOriginOrbitalArchitecture2005,morbidelliChaoticCaptureJupiters2005}, which explains the late rearrangement of the giant planets' orbits. Now, with thousands of multi-planet systems, we are finally able to compare observational data with results from global planetary formation models such as the Bern Model, refining and testing our planetary formation theories.

The classification of planetary system architectures serves multiple purposes: the effective comparison between systems (between observed and modelled systems, or systems from different catalogues), which helps validate or refute planetary formation theories; identifying general groups of planetary systems that shared formation processes, which helps define common evolutionary branches; improving exoplanet detection strategies by understanding typical planetary system structures, identifying potentially habitable systems because the system architecture influences planetary stability and composition which are critical factors for habitability; and finally predicting the presence of undetected planets with models such as the ones proposed in \cite{dietrichStatisticalDistributionFunction2024}. Planetary system architecture refers to the arrangement of planetary properties within a system, which can take various forms. Studies classify architectures based on orbital structure, which groups systems by planetary spacing or dynamical stability \cite{chambersHybridSymplecticIntegrator1999,lissauerARCHITECTUREDYNAMICSKEPLERS2011,laskarAMDstabilityClassificationPlanetary2017,stalportGlobalDynamicsArchitecture2022}, or by planetary mass and radius \cite{gilbertInformationTheoreticFramework2020,millhollandKeplerMultiplanetSystems2017,weissCaliforniaKeplerSurveyPeas2018}: some systems contain planets of similar sizes or masses, while others exhibit significant diversity \cite{millhollandKeplerMultiplanetSystems2017,mishraFrameworkArchitectureExoplanetary2023}. A global planetary formation model such as the Bern Model provides a wealth of data for such studies, particularly in defining architecture classes \cite{mishraFrameworkArchitectureExoplanetary2023,mishraFrameworkArchitectureExoplanetary2023a,emsenhuberPlanetaryPopulationSynthesis2023, davoultEarthlikePlanetsHosting2024} and predicting additional planets \cite{davoultEarthlikePlanetsHosting2024}.

\cite{mishraFrameworkArchitectureExoplanetary2023} define four distinct architecture classes based on the planetary mass distribution within systems of at least three planets:
\begin{itemize}
    \item Similar: All planets have similar masses,
    \item Ordered: Planetary masses tend to increase with distance from the star,
    \item Anti-Ordered: Planetary masses tend to decrease with distance from the star,
    \item Mixed: No clear trend in mass variation with distance from the star.
\end{itemize}
In \cite{mishraFrameworkArchitectureExoplanetary2023a}, these architecture classes are linked to different formation pathways. Similar systems tend to form in lower-mass protoplanetary disks associated with metal-poor stars, whereas the other three types arise in more massive disks around more metal-rich stars. The distinction between Ordered, Anti-Ordered, and Mixed architectures depends on dynamical interactions during system formation: the more of these interactions, the more likely the final architecture is to be Mixed, Anti-Ordered, or Ordered.

In \cite{davoultEarthlikePlanetsHosting2024}, we refine this framework using Principal Component Analysis (PCA) to study planetary mass distribution within systems. By considering the number of planets, the mass of the most massive planet, and PCA-derived properties, we define five architecture classes (Low-mass, Anti-Ordered, Ordered, Mixed, and n=1) inspired by the classes in \cite{mishraFrameworkArchitectureExoplanetary2023}. These classes, when correlated with the mass, radius, and orbital period of the innermost detected planet (IDP) and the host star’s mass, help identify systems most likely to host Earth-like planets (ELPs) among synthetic systems from the Bern Model. The results indicate that systems with Low-mass or n=1 architectures host the most ELPs. Additionally, the mass, radius, and period of the IDP are key indicators of ELP-hosting systems. Systems with a short-period, low-mass, or small-radius IDP, as well as those around intermediate-mass stars (G-type and late-K/early-M types), are more likely to host ELPs than systems around lower-mass late-M stars (see Tab.~\ref{tab:Davoult2024_summary} for a summary).

In \cite{Davoult2025}, these results are used to develop a model incorporating a random forest classifier trained to recognize ELP-hosting systems among synthetic systems from the Bern Model. This model was then applied to a list of 1567 observed systems with at least one known planet, identifying 44 systems as the most likely ELP hosts and thus high-priority targets for future studies.

In \cite{emsenhuberPlanetaryPopulationSynthesis2023}, we identify four architecture classes based on planetary mass-distance diagrams and the bulk composition of planets, considering five composition types. Instead of using a mathematical framework for classification, here, we visually inspect 1000 synthetic systems and define four primary architecture classes:
\begin{itemize}
    \item Class I: Compositionally ordered Earth and ice world systems,
    \item Class II: Migrated sub-Neptune systems,
    \item Class III: Mixed systems with low-mass and giant planets,
    \item Class IV: Dynamically active giant planet systems.
\end{itemize}
These architecture classes also appear to follow distinct formation pathways, a pattern already observed in \cite{mishraFrameworkArchitectureExoplanetary2023a}. We find that the primary factors influencing the classification are disk mass and disk lifetime, where Class I systems emerge from low-mass disks, Class II systems are associated with intermediate mass disks and longer disk lifetimes, and the giant hosting systems of Class III and IV form in the most massive disks with no obvious trend with the disk lifetime.\\

Synthetic planetary system populations computed with the Bern Model play a crucial role in studying planetary system architectures. These studies are essential for gaining a deeper understanding of planetary formation processes, enhancing comparisons between modelled and observed data, and providing predictions for future observations. By refining classification methods and linking architectures to formation pathways, these studies contribute significantly to the ongoing exploration of planetary system diversity and the search for habitable worlds.

\subsection{Conclusions}
Planet formation is a complex process because it involves many physical mechanisms that act concurrently on similar timescales, feeding strongly back on each other in a non-linear way \cite{Drazkowska2023ppvii, MordasiniBurn2024RvMG}. 

While specialized models for individual physical processes (like for example for planetary orbital migration or disk evolution) form the basis of planet formation theory, global models are needed to see their interactions and their impact on the finally forming planetary systems. It is also via global models predicting many directly observable quantities of (extrasolar) planets that theoretical models for specific mechanisms can be put to the observational test. In this way, global models form a bridge between theory and observations, and reflect the state of the field overall, giving constraints to the specialized models that are difficult to obtain otherwise.

One of the most comprehensive approaches to date to model planet formation and evolution in a global and end-to-end way is the Bern Model, which has been continuously developed for more than 20 years \cite{alibertModelsGiantPlanet2005}, and in particular during the NCCR PlanetS \cite{alibertFormationJupiterHybrid2018,emsenhuberNewGenerationPlanetary2021}. 

The model can be used to conduct planetary population syntheses \cite{benzPlanetPopulationSynthesis2014}, but also to study the impact of many different physical processes on the final outcome when embedded in a global model \cite{kaufmannInfluencePlanetesimalFragmentation2023,wederPopulationStudyMHD2023}.

Building upon earlier versions of the Bern Model \cite{alibertTheoreticalModelsPlanetary2013,mordasiniCharacterizationExoplanetsTheir2012}, the Generation III Bern Model was used to conduct the New Generation Planetary Population Synthesis (NGPPS) \cite{emsenhuberNewGenerationPlanetary2021a}. This large data set has been since used to study a multitude of important open questions in planet formation and evolution theory, like planet formation around host stars of different masses \cite{burnNewGenerationPlanetary2021}, the relation of inner low-mass planets and outer giants \cite{schleckerNewGenerationPlanetary2021}, the origin of the radius valley \cite{burnRadiusValleyMigrated2024}, system architectures and classifications \cite{mishraFrameworkArchitectureExoplanetary2023,schleckerNewGenerationPlanetary2021a}, or the peas-in-a-pod effect \cite{mishraNewGenerationPlanetary2021a}. Importantly, the data was also used to compare quantitatively to results of large observational surveys employing different methods \cite{viganSPHEREInfraredSurvey2021,Matuszewski2023} and to interpret observational results obtained with facilities with key NCCR PlanetS involvement, like CHEOPS \cite{eggerUnveilingInternalStructure2024} or high-precision radial velocity observations \cite{Unger2021,NGPPS7}, thereby forming an important connecting element in the NCCR.

Still, the Generation III model was built on classical underlying concepts like in particular viscous accretion disks without substructures and planetesimal accretion and giant impacts as the only solid accretion mechanism. In the chapter, we have therefore discussed several subsequent improvements to the Bern Model, like MHD wind-driven disks, planetesimal fragmentation, dust evolution and pebble accretion, structured disks, or an improved interior structure model.

While still a strong simplification of the actual planet formation process, these improvements make it in particular possible to directly simulate in one model the formation from sub-mm dust grains to gigayear-old fully-fledged planetary systems. 

Given the many observational facilities targeting extrasolar and Solar System planets that have recently started to observe or will do so in the near future, we can expect that the comparison of new observational constraints and future iterations of the Bern Model, but also other similar efforts (see \cite{burnPlanetaryPopulationSynthesis2024} for an overview) will make it possible to improve our understanding of the origin and evolution of planets.




\bibliographystyle{spphys_ed}
\bibliography{references}

\end{document}